\documentclass[twocolumn,aps,prl,raggedbottom,nobalancelastpage,amssymb,superscriptaddress,showpacs,10pt]{revtex4-1}

\usepackage{amsmath}
\usepackage{amssymb}
\usepackage{amsfonts}
\usepackage{dsfont}
\usepackage{graphicx}
\usepackage{bm}
\usepackage{color}
\usepackage{appendix}
\usepackage{epsfig}

\newcommand{\bra}[1]{\left<{#1}\right|}
\newcommand{\ket}[1]{\left|{#1}\right>}
\newcommand{\braket}[2]{\left<\left.{#1}\right|{#2}\right>}

\begin{document}
\title{Null Values and Quantum State Discrimination}

\author{Oded Zilberberg}
\affiliation{Department of Condensed Matter Physics, Weizmann Institute of Science, Rehovot 76100, Israel.} %
\author{Alessandro Romito}
\affiliation{\mbox{Dahlem Center for Complex Quantum Systems and Fachbereich Physik, Freie Universit\"at Berlin, 14195 Berlin, Germany}}
\author{David J. Starling}
\affiliation{Department of Physics and Astronomy, University of Rochester, Rochester, New York 14627, USA.}
\affiliation{Division of Science, Pennsylvania State University, Hazleton, Pennsylvania 18202, USA.}
\author{Gregory A. Howland}
\affiliation{Department of Physics and Astronomy, University of Rochester, Rochester, New York 14627, USA.} %
\author{Curtis J. Broadbent}
\affiliation{Department of Physics and Astronomy, University of Rochester, Rochester, New York 14627, USA.} %
\affiliation{Rochester Theory Center, University of Rochester, Rochester, New York 14627, USA.} %
\author{John C. Howell}
\affiliation{Department of Condensed Matter Physics, Weizmann Institute of Science, Rehovot 76100, Israel.}
\affiliation{Department of Physics and Astronomy, University of Rochester, Rochester, New York 14627, USA.} %
\author{Yuval Gefen}
\affiliation{Department of Condensed Matter Physics, Weizmann Institute of Science, Rehovot 76100, Israel.}%

\begin{abstract}
 We present a measurement protocol for discriminating between two different quantum states of a qubit with high fidelity. The protocol, called \textit{null value}, is comprised of a projective measurement performed on the system with a small probability (also known as partial-collapse), followed by a tuned postselection. We report on an optical experimental implementation of the scheme. We show that our protocol leads to an amplified signal-to-noise ratio (as compared with a straightforward strong measurement) when discerning between the two quantum states.
\end{abstract}

\maketitle

The notion of ``measurement'' was part of the early framework of quantum mechanics.
Since early developments, the discord between information acquisition on the system and the corresponding disturbance of the system's state became clear.
The contest of obtaining information while keeping minimal disturbance is still an active and vibrant field of study that has branched off into many subtopics. Of note and of great practical interest in quantum information processing is the study of quantum state discrimination~\cite{Helstrom,Loss:1998,Chefles_review,Wiseman:2010}.
The ability to optimally discriminate between nonorthogonal quantum states depends on the fidelity of the measurement apparatus and on the amount of prior knowledge one has on the states between which he wants to distinguish.

Here we introduce a novel procedure to enhance the discrimination fidelity between two quantum states. Our procedure introduces the notion of quantum measurements with postselection in the field of quantum information processing. Our two-step measurement protocol is related, but differs from, the celebrated weak value (WV) measurement protocol, where postselected quantum measurements were first introduced~\cite{Aharonov:1988aa}.
We choose to demonstrate our new approach by focusing on a specific discrimination problem. In conjunction with our theoretical analysis, we report on experimental results involving classical light, which demonstrate the practicality of our measurement protocol, denoted ``null value'' (NV) measurement protocol.

In the original works on quantum state discrimination, the observer is handed a single copy of the state to be discriminated, which may be either one of the \textit{a priori} known pure states $\ket{A}$ and $\ket{B}$.
Well adapted to this task is the approach known as \textit{minimum error state discrimination}~\cite{Helstrom}, for which it was shown that the minimum error is obtained by optimizing the axis of a standard two-outcome measurement.
A second approach is the \textit{unambiguous state discrimination}~\cite{Ivanovic:1987,Dieks:1988,Peres:1988} where the measurement produces either an error-free or an inconclusive result; i.e., the measurement apparatus is oriented such that it has three outcomes -- the state is either $A$, $B$, or unknown.

Developments on the original works led to many variants of state discrimination, such as discrimination between two \textit{a priori} unknown pure states~\cite{Massar:1995,Derka:1998}, as well as discrimination between mixed states~\cite{Osaki:1996,Higgins:2009}.
   Further works (see e.g., Refs.~\cite{Walgate:2000,Acin:2005}) introduced also the notion of multicopy state discrimination (employing a number of copies of the state to be discriminated). For such schemes, the notion of individual vs. collective measurements was introduced depending on whether the strategy consists of individual measurements each of which performed separately on a single copy, or a single measurement which is performed on all the copies as a whole. Notably, however, in most standard measurement procedures one performs individual measurements on $N$ single copies. Thus, it is necessary to define statistical tests to quantify the fidelity of the discrimination~\cite{Engel:2004}.

In the present scheme, we study a specific variant of the quantum state discrimination problem: the observer prepares a device (a protocol) that should discriminate whether the provided state is equal to the known state $\ket{\psi_0}$ or not, i.e. is some other nearby state $\ket{\psi_\delta}$.
Noting the context of earlier works on state discrimination, our variant applies to both single-copy and many-copy analyses \cite{supplementary}. In the former, due to the \textit{a priori} unknown orientation of $\ket{\psi_\delta}$, a minimum error state discrimination is underconstrained. Additionally, an unambiguous state discrimination is impossible as the unknown state would generate both erroneous and inconclusive results. Such a case, dubbed ``intermediate discrimination scheme'' has been treated for discrimination between two different states (see e.g., Refs.~\cite{Chefles:1998b,Miroslav:2003,Wittmann:2010}). 

We present our analysis henceforth for discrimination between two-level states (qubits). Assuming that the probability distribution of $\ket{\psi_\delta}$ is uniform on the Bloch sphere in some area around $\ket{\psi_0}$, we have analyzed the single-copy minimum error and the single-copy intermediate schemes vis-a-vis our discrimination problem~\cite{supplementary}. For the former, we obtain that, regardless of the area of the distribution, minimum error is obtained for a standard measurement in the direction orthogonal to $\ket{\psi_0}$. For the latter, we recall that a three-outcome measurement on a qubit can be realized by two-consecutive measurements \cite{Chefles_review}. Their optimal orientations depend on the area of the distribution. For discrimination between nearby states ($\ket{\psi_0}$ or not $\ket{\psi_0}$), the optimal orientations of both measurements are nearly orthogonal to $\ket{\psi_0}$. By contrast, when the probability distribution of $\ket{\psi_\delta}$ covers the entire Bloch sphere the first of the two measurements is oriented in the direction of $\ket{\psi_0}$ itself.

Alas, a single-copy approach is unfit for most experimental situations due to measurement device imperfections and noise. One then resorts to a multicopy approach. Here one considers a statistical test (signal-to-noise ratio, SNR) that, given $N$ replicas of the state, would result in a discrimination outcome ($\ket{\psi_0}$ or not $\ket{\psi_0}$) with some given fidelity. Below we define such SNRs and employ them to compare a multicopy version of both the minimum error scheme and the intermediate scheme, focusing on a discrimination between nearby states. For the intermediate scheme we consider a correlated signal, dubbed \textit{null value signal}, for a reason that will be made clear below.
The SNR obtained by a NV signal proves to be higher than that obtained by single von Neumann measurements.
We further show that the analysis in terms of optimal SNR fully agrees with a minimization of error probability in the single-copy cases~\cite{supplementary}.

Let us begin with analyzing the SNR of the discrimination, achieved through individual standard strong measurement, $M_s$, on $N$ copies of a qubit. In this benchmark case, the occupation of the state $\ket{M}$, defined by polar angle $\theta_M$, is measured~\cite{polar}. The probabilities to detect the qubit states $\ket{ \psi_0 }$ and $\ket{ \psi_\delta }$ with polar angles 0 and $\delta$ in $\ket{M}$ in any single attempt are $P(M_{s,0})=|\braket{M}{\psi_0}|^2$ and $P(M_{s,\delta})=|\braket{M}{\psi_\delta}|^2$, respectively. We define a statistical measure to be the difference between the number of positive detections 
\begin{equation} S_{\text{std}}=N\left|P(M_{s,\delta})-P(M_{s,0})\right|\cong\left|N_{s,\delta}-N_{s,0}\right|\, ,
	\label{signal_simp}
\end{equation}
where the right-hand side is the measured estimator. The signal is a function of two variables $S_{\text{std}}(N_{s,\delta},N_{s,0})$. The uncertainty in the signal is then given by
\begin{align}
\Delta S_{\text{std}}=\sqrt{\left(\frac{\partial S_{\text{std}}}{\partial N_{s,\delta}}\right)^2\Delta N_{s,\delta}^2 +\left(\frac{\partial S_{\text{std}}}{\partial N_{s,0}}\right)^2\Delta N_{s,0}^2}\, .
	\label{noise_simp}
\end{align}
We assume Poissonian noise (which is dominant for coherent light experiments discussed below), i.e., $\Delta N_{s,\delta}^2=N_{s,\delta}$ and $\Delta N_{s,0}^2=N_{s,0}$. Thus, $\Delta S_{\text{std}}=\sqrt{N_{s,\delta}+N_{s,0}}$, and the obtained SNR is
\begin{equation}
	\mathrm{SNR}_{\text{std}}=\frac{S_{\text{std}}}{\Delta S_{\text{std}}}\approx \sqrt{2}|\sin[\theta_M]|\delta \sqrt{N} \,, 
	\label{SNR_simp}
\end{equation}
where the approximation is for $\delta \ll 1$. Indeed, in this approach the maximal SNR is obtained when the measurement orientation, $\ket{M}$, is orthogonal to $\ket{\psi_0}$. This corresponds to the optimal measurement orientation obtained by the single-copy analysis.

\begin{figure}
\includegraphics[width=0.8\columnwidth]{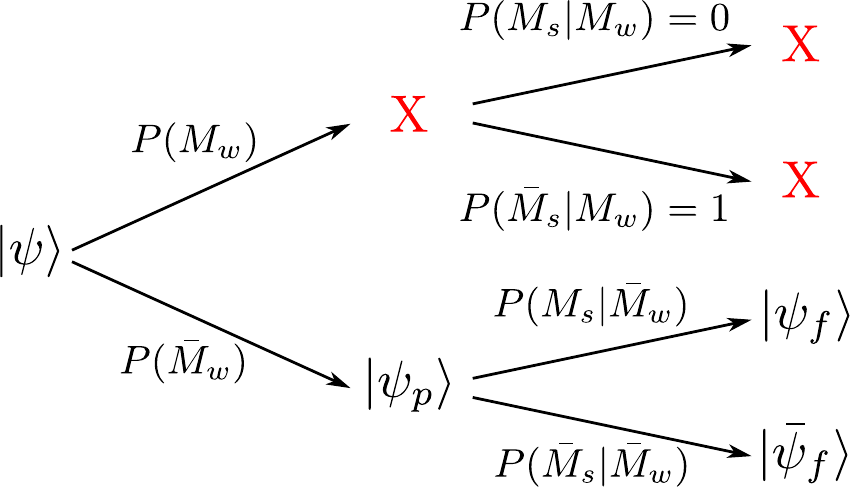}
\caption{\label{Fig:1}
A tree diagram  of the qubit state evolution under subsequent partial-collapse measurements; the respective  probabilities are indicated: $P(M_{w})$ [$P(\bar{M}_{w})$] is the probability that the detector ``clicks'' [no ``click''] upon the first measurement. If it does ``click'', the system is destroyed, hence there are no clicks upon further measurements [this is marked by a (red) $\mbox{X}$]. Note that following $P(\bar{M}_{w})$ (null detection of the qubit), the back action rotates $\ket{\psi}$ into $\ket{\psi_p}$.}
\end{figure}

Turning to the multicopy intermediate discrimination, we define a SNR by constructing a correlated outcome out of the three-outcome measurement. 
Recall that such a measurement is implemented by measuring the qubit state twice (cf. Fig.~\ref{Fig:1}) \cite{supplementary}.
The first measurement $M_w$ is a strong (projective) measurement which is performed on the system with small probability.
Here the basis states $\{\ket{\bar{M}},\ket{M}\}$ are measured with probabilities $\{p_0,p_1\}$, respectively.
For simplicity, hereafter, we assume that only the state $\ket{M}$ is measured with probability $p_1=p$ and $p_0=0$.
If the detector ``clicks'' (the measurement outcome is positive), the qubit state is destroyed. Very importantly, having a ``null outcome'' (no click) still results in a back action on the system. We refer to this stage of the measurement process as ``partial-collapse'' \cite{Korotkov:2007}.
Subsequently the qubit state is (strongly) measured a second time (postselected), $M_s$, to be in the state $\ket{\psi_f}$ (click) or $\ket{\bar{\psi}_f}$ (no click), where $\ket{\psi_f}$ has a polar angle of $\theta_f$. We propose to discriminate between the two possible initial qubit states by individual application of this measurement protocol on $N$ copies of $\ket{\psi_0}$ and $\ket{\psi_\delta}$. Motivated by WVs, the compared observables are the countercausal conditional outcome of [having a click the first time conditional to \textit{not} having a click the second time], denoted by $P(M_{w,0}|\bar{M}_{s,0})$ and $P(M_{w,\delta}|\bar{M}_{s,\delta})$, respectively. Events in which the qubit is measured strongly (in the second measurement), $M_s$, are discarded. In other words, we define our signal to be
\begin{equation}
S_{\text{NV}}\equiv N\left| P(M_{w,\delta}|\bar{M}_{s,\delta})-P(M_{w,0}|\bar{M}_{s,0})\right|\,.
\end{equation}
Note that this procedure can also be written as a statistical correlation between outcomes of a positive-operator valued measure (POVM)~\cite{supplementary}.

Our protocol takes advantage of the statistical correlations between the partial-collapse and strong measurements. To shed some light on its outcome we calculate explicitly the conditional probabilities following the measurement procedure sketched in Fig.~\ref{Fig:1}.  For example, if the first measurement results in a ``click'' the system's state is destroyed and any subsequent measurement on the system results in a null result.  This represents a classical correlation between the two measurements. By contrast, $P(\bar{M}_s|\bar{M}_w)$ embeds nontrivial quantum correlations \cite{noclick}.
Using Bayes theorem, we can write $ P(M_{w,\delta}|\bar{M}_{s,\delta}) \cong N_{w,\delta}/ (N_{w,\delta}+N_{p,\delta})$,
where we used the measured estimator for the conditional probability, namely, we denoted $N_{w,\delta}\cong N P(M_{w,\delta})$ as the number of clicks in the (first) partial-collapse measurement and $N_{p,\delta}\cong N P(\bar{M}_{w,\delta}) P(\bar{M}_{s,\delta}|\bar{M}_{w,\delta})$ as the number of no-clicks in the (second) postselection~\cite{supplementary}. This finally leads to the measured signal
\begin{equation}
	S_{\text{NV}}\cong N\left|\frac{N_{w,\delta}}{ N_{w,\delta}+N_{p,\delta}}-\frac{N_{w,0}}{ N_{w,0}+N_{p,0}}\right|\,.
	\label{signal_nullwv}
\end{equation}
In complete analogy with the case of a single strong measurement, the signal is now a function of four variables $S_{\text{NV}}(N_{w,\delta},N_{p,\delta},N_{w,0},N_{p,0})$, and we can define the uncertainty, $\Delta S_{\text{NV}}$, in the statistical test [cf.~Eq.~\eqref{noise_simp}] \cite{supplementary}.

We focus on obtaining a large $\text{SNR}_{\text{NV}}=S_{\text{NV}}/\Delta S_{\text{NV}}$ for discriminating between the two states. It depends on the choice of the measurement orientations, $\ket{M}$ and $\ket{\psi_f}$.
We propose to perform a first measurement that will have a back action on both states $\ket{\psi_0}$ and $\ket{\psi_\delta}$ but is nearby the optimal orientation of the single measurement case, i.e. taking $\theta_M=\pi/2+\Delta_M$. We propose two possible measurement schemes for obtaining a large $\text{SNR}_{\text{NV}}$. In the first scheme we choose the postselection such that the reference state satisfies $|\braket{\bar{\psi}_f}{\psi_0}|^2=0$. This means that the reference state $\ket{\psi_0}$ would have always clicked in the second measurement had it not been first measured by the partial-collapse. We call this \textit{scheme \textbf{A}}. Alternatively, in \textit{scheme \textbf{B}}, we choose the postselection such that $|\braket{\bar{\psi}_f}{\psi_{0,p}}|^2=0$: the null outcome rotates the reference state and it always clicks in the second measurement. For both schemes, we obtain
\begin{equation}
\textrm{SNR}_{\text{NV}} (N)\sim \frac{\sin^2[\delta]}{\sin[\Delta_M+\delta]\sqrt{p}}\sqrt{N}\,,
\end{equation}
which becomes large for $p\rightarrow 0$ (weak partial-collapse) \cite{supplementary}. This is because the condition $N_{w,0}\gg N_{p,0}$ is satisfied vis-a-vis the NV signal of the reference state. Varying $\delta$ such that $|\braket{\psi_\delta}{\psi_0}|^2$ decreases corresponds to a decrease of $N_{w,\delta}$ and an increase of $N_{p,\delta}$. A large $\text{SNR}_{\text{NV}}$ is obtained when $P(M_{w,\delta}|\bar{M}_{s,\delta})$ crosses to a regime where $N_{w,\delta}\leq N_{p,\delta}$. This happens first with scheme \textbf{\textit{A}}. Hence, scheme \textbf{\textit{A}} produces a larger $\text{SNR}_{\text{NV}}$ for smaller $\delta$; scheme \textbf{\textit{B}} leads to far larger $\text{SNR}_{\text{NV}}$ for larger $\delta$. Note, also, that taking the partial-collapse measurement to be more orthogonal to $\ket{\psi_0}$ ($\Delta_M$) increases the SNR for $\delta\ll 1$.
 
The postselection measurement orientations, which produce the high SNR, coincide with those obtained in the single-copy analysis, i.e. $|\braket{M}{\psi_0}|\sim 0$, $p\ll 1$, for $\delta\ll 1$ \cite{supplementary}. This suggests that though the spirit of the present multicopy analysis is quite different from the single-copy analysis, both analyses give similar guidance for optimally discriminating between nonorthogonal states. We reiterate, however, that (as compared with the single-copy approach) the statistical SNR approach (based on NV) is better suited to most experimental settings in which noise and experimental imperfections are present.

\begin{figure}
  \begin{center}
    \includegraphics[width=\columnwidth]{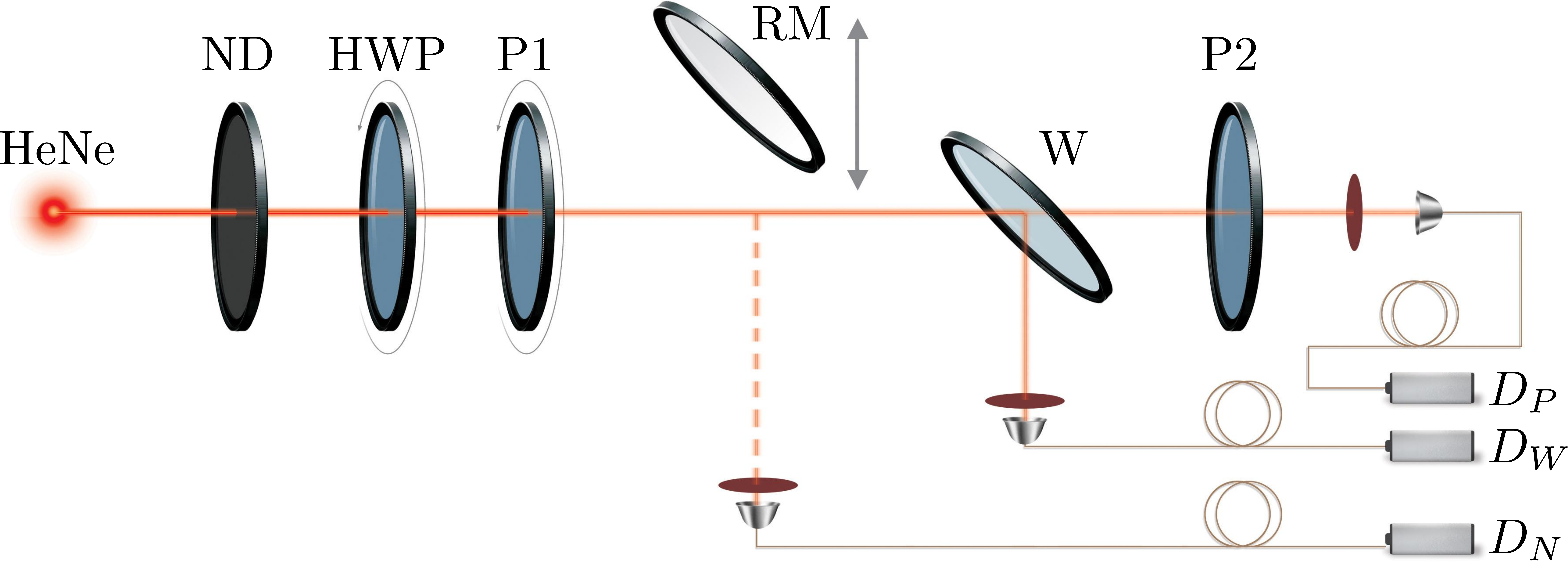}
  \end{center}
  \caption{\label{Fig:2}
  A sketch of the experimental apparatus. Single spatial mode light from a helium-neon laser (HeNe) passes through a neutral density filter (ND) followed by a half-wave plate (HWP) and polarizer (P1) to prepare the initial state. During data acquisition, the HWP is used to maintain a constant photon flux which is measured using a removable mirror (RM). A glass window (W) weakly reflects vertically polarized light. Photons that pass through the window are then projected onto a linear polarization state with a second polarizer (P2). The photons in each spatial mode are passed through colored glass filters to block background, collected via a multimode fiber and sent to single photon counting modules ($D_N$, $D_W$ and $D_P$).}
\end{figure}

We measure the NV signal and its amplified SNR using an optical technique sketched in Fig.~\ref{Fig:2}. Here, the qubits are replaced by photons from a dramatically attenuated coherent beam, and the measurement device consists of polarization optics and single-photon detectors. We encode the states in the polarization degree of freedom by passing the beam through a polarizer (P1), giving $\ket{\psi_\delta}= \cos[\delta-\Delta_M] \ket{0} + \sin[\delta-\Delta_M]\ket{1}$, where $\{\ket{0},\ket{1}\}$ correspond to the horizontal and vertical polarization states, respectively. We perform a (weak) partial-collapse measurement by sending the photons through a glass window (W) set at the Brewster angle. The window therefore weakly reflects vertically polarized light, with probability $p=0.15$, and passes horizontal light with near unit probability. We set the second polarizer (P2) in the transmitted arm to strongly project the photon into the state $\ket{\bar{\psi}_f}$ which is represented by scheme \textbf{\textit{A}} or \textbf{\textit{B}}, as desired \cite{rotated}. From the resulting photon detections we obtain the values of $N_{w,\delta}$, $N_{p,\delta}$, $N_{w,0}$ and $N_{p,0}$ and their variances \cite{supplementary}.

\begin{figure}
  \begin{center}
    \includegraphics[width=\columnwidth]{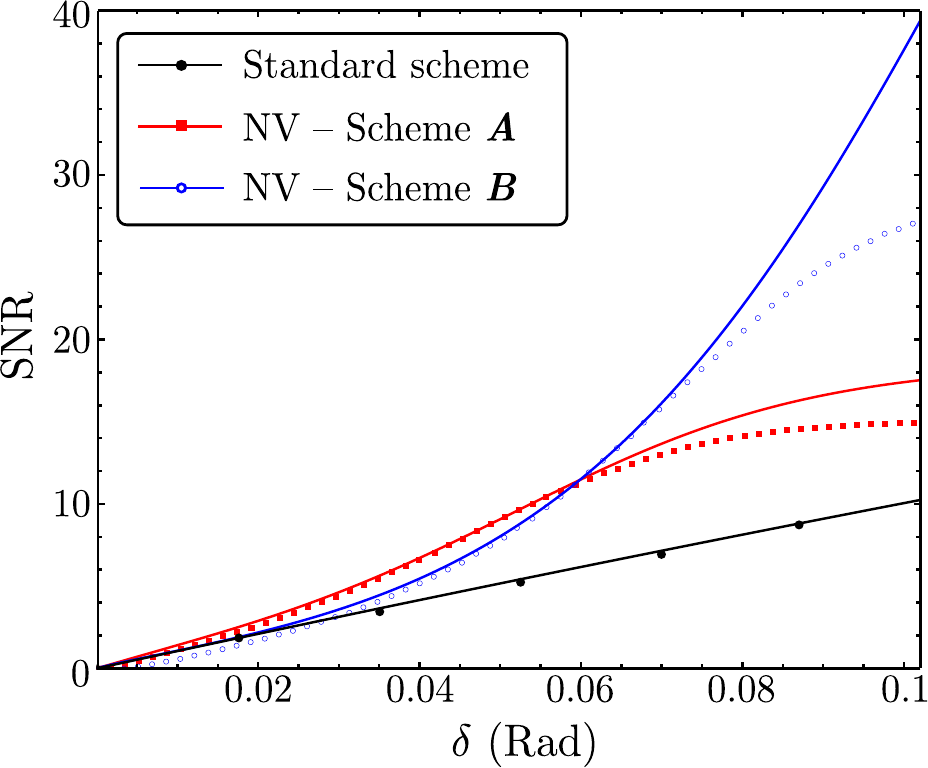}
  \end{center}
  \caption{\label{Fig:3}
  A graph of the theoretical and experimental SNR obtained for different measurement schemes. Scheme \textbf{\textit{A}} (red squares) and \textbf{\textit{B}} (blue hollow circles) correspond to the null value technique ($\text{SNR}_{\text{NV}}$). The parameter $\delta$ denotes the distance between the measured and the reference state; it is varied by changing the angles for the input polarizer P1. For a given P2 and W (cf.~Fig.~\ref{Fig:2}) the reference state is determined by finding P1 for which $|\braket{\bar{\psi}_f}{\psi_0}|^2$, $|\braket{\bar{\psi}_f}{\psi_p}|^2$ is minimal for schemes \textbf{\textit{A}}, \textbf{\textit{B}}, respectively. The standard scheme (black circles) is that defined by Eq.~(\ref{SNR_simp}), and is represented by a single polarizer with no weak measurement. Dots correspond to calculations from data and lines correspond to the theoretical predictions. Each scheme used approximately the same number of photons, with $N \approx 11250$ per measurement.}
\end{figure}

We consider schemes \textbf{\textit{A}} and \textbf{\textit{B}} for $\Delta_M = 0.1$ rad and plot the results in Fig.~\ref{Fig:3}. We find that, for  scheme \textbf{\textit{A}}, we can discriminate between the two states with a higher SNR than the standard scheme nearly over the whole range of angles considered. Similarly, while the SNR of the standard technique almost coincides with that of scheme \textbf{\textit{B}} for small angles, we see that the sensitivity of the two schemes diverges quickly for larger angles; in this regime ($\delta \approx \Delta_M$), the NV scheme \textbf{\textit{B}} is significantly better. 
The discrepancy between theory and experiment is due to a small amount of ellipticity incurred from the glass window not included in the theory plot \cite{supplementary}. 

The described NV procedure leading to large SNR is based on the conditional outcome of a quantum measurement. As such, it resembles the well-established protocol of WV measurement~\cite{Aharonov:1988aa}.
The WV protocol consists of weakly measuring an operator $\hat{A}$ of a \textit{system} prepared in an initial state $\ket{i}$ by weakly coupling it to a \textit{detector}. The detector output is kept only if the system is eventually measured to be in a chosen final state, $\ket{\psi_f}$---\emph{postselection}. The obtained conditional average of $\hat{A}$, $\bra{f}\hat{A}\ket{i}/\braket{f}{i}$, is named \textit{weak value}, and can be anomalously large~\cite{Aharonov:1988aa}. This property has been exploited for amplifying small signals both in  quantum optics~\cite{Hosten:2008,Dixon:2009,Starling:2009,Brunner:2010,Starling:2010b} and in solid state physics~\cite{Zilberberg:2011}. 
It is important to stress that the NV protocol is different from the WV protocol. The former makes use of a partial-collapse measurement of the operator $\hat{A}$, in which the system experiences back action only for a subset of all possible measurement outcomes, while a strong projection takes place for the remaining outcomes. This is not a weak measurement, which is used by the WV protocol. The obtained conditional average of $\hat{A}$ is now the NV, $(1/p)P(M_w|\bar{M}_s)=\bra{i}\hat{A}\ket{i}/P(\bar{M}_s)$. It is quantitatively different from the WV even when $p$ is explicitly ``weak'' \cite{proceeding}. Moreover, while a large WV leads to an amplification of the SNR for systems where the noise is dominated by an external (technical) component~\cite{Starling:2009,Zilberberg:2011},
the method presented here leads to high fidelity discrimination between quantum states on the background of quantum fluctuations. 

 In conclusion we have presented here a new protocol based on a  partial-collapse measurement followed by a tuned postselection. Our protocol enables one to discern between quantum states with better accuracy than a standard measurement would allow. By contrast to earlier protocols~\cite{Chefles_review} tuned to discriminate between two prescribed states, the present one facilitates the study of an amplified SNR for a wide range of possible polarizations of one of the states, which is not \textit{a priori} known. We have demonstrated the feasibility and effectiveness of our protocol by employing an optical setup for discriminating between different polarization states of light. Notably, our present approach is based on a statistical analysis, which makes it particularly suitable for experiments.

We would like to thank N.~Katz, V. Giovannetti, C. Bruder, G. Str\"{u}bi, and A.~Ustinov for fruitful discussions, and N.~Gontmakher for the illustration of the setup. This work was supported by GIF, Einstein Minerva Center, ISF, Minerva Foundation, Israel-Korea MOST grant, and EU GEOMDISS. J.~C.~H.~was supported by Army Research Office grants W911 NF-12-1-0263 and W911-NF-09-0-01417. C.~J.~B.~acknowledges support from ARO W911NF-09-1-0385 and NSF PHY-1203931.


\newpage
\begin{center}
\textbf{\large SUPPLEMENTAL MATERIAL}
\end{center}
\appendix
In this Supplemental Material, we provide additional technical details of our analysis and experimental methods. Section \ref{nvwv} discusses the importance of the negative outcome postselection (no click). In Section \ref{povm}, we provide a translation between the conditional probability formalism, used in the main text, and positive-operator valued measure (POVM) language. Section \ref{single_copy} describes the technical details of the single-copy minimum error scheme and the single-copy intermediate scheme vis-a-vis our discrimination problem. Section \ref{multi_copy} presents additional technical details of the multi-copy minimum error scheme (performed by individual standard strong measurements) and the multi-copy intermediate scheme (null values). Finally, in Section \ref{experimental_methods}, we provide additional details on our experimental methods with a focus on sources of loss. 

\section{I. The importance of the negative outcome (no click) postselection}
\label{nvwv}
In both the weak value (WV) and null value (NV) measurement protocols there are two consecutive measurements. Generally, using binary detectors (``click'' or ``no click''), we would have four possible outcomes at any given attempt:  (``click'', ``click''), (``click'', ``no click''), (``no click'', ``click''), (``no click'', ``no click''), where the notation here is (first measurement outcome, second measurement outcome). Taking the outcome of a first weak measurement conditional to a strong second measurement (i.e., how the outcome of the first measurement is correlated with the given outcome of the second measurement) would be proportional to the (real or imaginary part of the) standard weak value (see e.g., Ref.~\cite{Romito:2008}).

A partial-collapse measurement, employed in the NV protocol, is a measurement for which a ``click'' destroys the system. Hence, in our scheme (``click'', ``click'') is not allowed, and there are only three possible outcomes: (``click'', ``no click''), (``no click'', ``click''), (``no click'', ``no click''). Note that having ``no click'' in the partial-collapse measurement still inflicts back action on the measured system.
For the reasoning above the standard correlation between a ``click'' in the first measurement conditional to a ``click'' in the second measurement is trivially zero. By contrast, the correlation between a ``click'' in the first measurement conditional to ``no click'' in the second measurement is not trivial, which is the motivation for our \textit{null} value scheme.

The experimental setup consists of an exact mapping of this idea. The crucial point here, though, is to note what would be considered to be ``click'' and ``no click'' in either the first measurement or the second one. To be specific, impinging upon the window ($W$) (cf. Fig.2), a photon can either be deflected to a photon counter $D_w$ (``click''), or pass through the window (``no click''). A photon that has passed through the window reaches the second polarizer $P_2$, where the second measurement takes place. It can either be absorbed by $P_2$ (``click'' in the second measurement) or pass through it to the photon counter $D_p$. Thus, counting a photon in $D_p$ will correspond to ``no click'' in the second measurement.

\section{II. POVM formalism}
\label{povm}
We provide here a description of the procedure leading to null values in terms of POVMs~\cite{Helstrom}. The procedure consists of two steps.
First, the state is measured using a partial-collapse measurement
\begin{align}
\label{K1}
K_1 &= \sqrt{p_0} \ket{\bar{M}}\bra{\bar{M}} + \sqrt{p_1} \ket{M}\bra{M}\,,\\
K_0 &= \sqrt{1-p_0} \ket{\bar{M}}\bra{\bar{M}} + \sqrt{1-p_1} \ket{M}\bra{M}\,,
\label{K0}
\end{align}
where $p_0,p_1$ are the probabilities for the states $\ket{\bar{M}},\ket{M}$ to ``click'', respectively. Note that in the main text we have taken $p_0=0,p_1=p$. Following the partial-collapse step, the state is  measured with a projective measurement $\{\ket{\psi_f}\bra{\psi_f},\ket{\bar{\psi}_f}\bra{\bar{\psi}_f}\}$. Consequently, we can represent the sequence of measurements with the following POVM elements,
\begin{align}
\hat{\Pi}_1 &= K_1^\dagger K_1 \,,\\
\hat{\Pi}_2 &= K_0^\dagger \Pi_f K_0 \,,\\
\hat{\Pi}_? &= K_0^\dagger \Pi_{\bar{f}} K_0\,,
\end{align}
where $\Pi_f=\ket{\psi_f}\bra{\psi_f}$ and $\Pi_{\bar{f}}=\ket{\bar{\psi}_f}\bra{\bar{\psi}_f}$. Indeed this set forms a POVM set $\hat{\Pi}_1+\hat{\Pi}_2+\hat{\Pi}_? = \mathds{1}$. Notably, unambiguous and intermediate state discriminations are performed by tuning such a three-component POVM set~\cite{Ivanovic:1987,Dieks:1988,Peres:1988}. There are several similar optical implementations of such three-component POVM (see e.g. Ref.~\cite{Wiseman:2010} and references therein).

The measurement probabilities in the manuscript can be expressed in terms of these POVM operators,
\begin{align}
P(M_{w,\delta}) = \bra{\psi_\delta}\hat{\Pi}_1\ket{\psi_\delta}\,,\\
P(M_{s,\delta} , \bar{M}_{w,\delta}) = \bra{\psi_\delta}\hat{\Pi}_2\ket{\psi_\delta}\,,\\
P(\bar{M}_{s,\delta} , \bar{M}_{w,\delta}) = \bra{\psi_\delta}\hat{\Pi}_?\ket{\psi_\delta}\,,
\end{align}
where we wrote here the joint probabilities $P(M_{s,\delta} , \bar{M}_{w,\delta})=P(\bar{M}_{w,\delta})P(M_{s,\delta} | \bar{M}_{w,\delta})$ and $P(M_{s,\delta} , \bar{M}_{w,\delta})=P(\bar{M}_{w,\delta})P(M_{s,\delta} | \bar{M}_{w,\delta})$.

Additionally, in Table~\ref{table1} we provide a translation of the POVM operators to ``clicks'' at different components in the experimental setup.
\begin{table}[htb]
\begin{tabular}{|c|c|c|c|}
 \hline
  POVM           & $D_W$    & $P_2$ & $D_P$  \\
  \hline
  $\hat{\Pi}_1$ & 1 & 0 & 0\\
  $\hat{\Pi}_2$ & 0 & 1 & 0\\
  $\hat{\Pi}_?$ & 0 & 0 & 1\\
  \hline
\end{tabular}
\caption{Translation of the POVM language to ``clicks'' in the experimental setup. A ``click'' is denoted by $1$ and ``no click'' by $0$. Note that while a ``click'' in the photon counters $D_W$ and $D_P$ is easily understood, a ``click'' in $P_2$ corresponds to a photon being absorbed by the polarizer. Hence, a ``no click'' at the postselection stage corresponds to a photon not being absorbed in $P_2$. } \label{table1}
\end{table}

The NV is composed of a correlation between the different POVM components but cannot be written directly as a POVM.

\section{III. Single-copy state discrimination}
\label{single_copy}
We study a branch of the quantum state discrimination problem where the observer prepares a device (a protocol) that should discriminate whether the provided state is equal to the known state $\ket{\psi_0}$ or not, i.e. some other state $\ket{\psi_\delta}$. We study two cases:

\subsection{A. Minimum error scheme}
 Discrimination using a single strong von Neumann measurement generates erroneous results with the probability 
 \begin{align}
 p_{\rm err}=1/2[ P( \psi_0 | \psi_\delta) + P( \psi_\delta | \psi_0 ) ]\,,
 \end{align}
  where $P( \psi_\delta | \psi_0 )$ is the probability of erroneously declaring the state to be $\ket{\psi_\delta}$ when it was in fact $\ket{\psi_0}$ and $P( \psi_0 | \psi_\delta)$ is the probability of the converse error.

\begin{figure}
  \begin{center}
    \includegraphics[width=0.5\columnwidth]{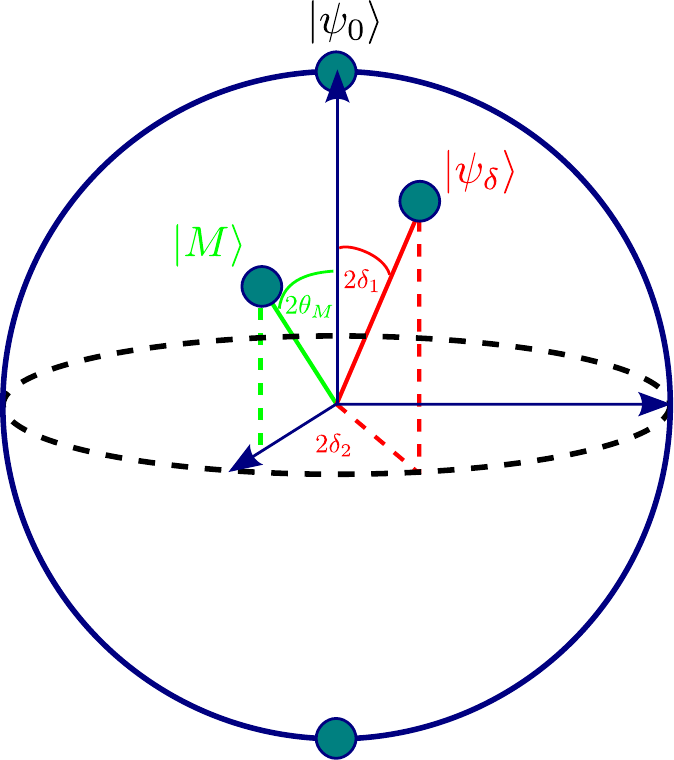}
  \end{center}
  \caption{\label{bloch1}
  A Bloch sphere visualization of the chosen gauge for the analysis of the single von Neumann measurement case.}
\end{figure}

 Let us choose a gauge such that $\ket{\psi_0}=\ket{0}$, the measurement orientation is $\ket{M}=\cos[\theta_M] \ket{0} + \sin[\theta_M]\ket{1}$, and the alternate state is $\ket{\psi_\delta}=\cos[\delta_1] \ket{0} + \sin[\delta_1]e^{i\delta_2}\ket{1}$ (see Fig.~\ref{bloch1}). If a ``click'' is obtained, we declare that the state $\ket{\psi_\delta}$ was measured. Hence,
\begin{align}
	 P( \psi_\delta | \psi_0 )=&\cos^2\left[\theta_M\right]\, ,\\
	 P( \psi_0 | \psi_\delta)=&\frac{1}{2}\Big(1-\cos[2\theta_M]\cos[2\delta_1]\nonumber\\
	 &-\sin[2\theta_M]\sin[2\delta_1]\cos\left[\delta_2\right]\Big)\, .
\end{align}

 The probability of error, $p_{\rm err}$, depends on the orientation of $\ket{\psi_\delta}$ such that one cannot find an optimal measurement orientation. Nevertheless, assuming a probability distribution out of which $\ket{\psi_\delta}$ is chosen, enables some optimization in the choice of the measurement direction.


We can assume, for example, that $\ket{\psi_\delta}$ is uniformly distributed on the Bloch sphere in some area around $\ket{\psi_0}$. Taking the mean over this distribution results in
\begin{align}
\label{mean_std}
	\overline{P( \psi_0 | \psi_\delta)}& =\frac{1}{S(\Delta)}\int_0^\Delta\;d\delta_1\int_0^{\pi}\;d\delta_2\;P(\psi_0 | \psi_\delta)\sin[2\delta_1]\nonumber\\
	&=\frac{1}{2}\left(1-\cos[2\theta_M]\cos^2\left[\Delta\right]\right)\,,
\end{align}
where $S(\Delta)=\int_0^\Delta\;d\delta_1\int_0^{\pi}\;d\delta_2\;\sin[2\delta_1]$. Averaging over the entire Bloch sphere amounts to taking $\Delta\rightarrow \pi/2$. Thus, the optimal direction to measure (minimizing $p_{\rm err}$) would be in the direction orthogonal to $\ket{\psi_0}$, i.e. taking $\theta_M=\pi/2$. This optimal measurement direction is the same as that obtained in the main text using the SNR analysis.

\subsection{B. Intermediate scheme (Three-component POVM)}
 Our type of intermediate discrimination using a three-component POVM generates both inconclusive an erroneous results~\cite{Chefles:1998b,Miroslav:2003,Wittmann:2010}. The probability for inconclusive results is 
 \begin{align}
 p_{\rm inc}=(1/2)[ \bra{\psi_0}\hat{\Pi}_?\ket{\psi_0} + \bra{\psi_\delta}\hat{\Pi}_?\ket{\psi_\delta}]\,.
 \end{align}
  The probability of error in this case is 
  \begin{align}
  \label{perr3povm}
  \bar{p}_{\rm err}=\frac{1}{2}\frac{ \bra{\psi_0}\hat{\Pi}_2\ket{\psi_0} + \bra{\psi_\delta}\hat{\Pi}_1\ket{\psi_\delta}}{1 - p_{\rm inc}}\,, 
  \end{align}
   where a ``click'' in $\hat{\Pi}_1$, $\hat{\Pi}_2$ is erroneously interpreted as the state being $\psi_0$, $\psi_\delta$, respectively.

\begin{figure}
  \begin{center}
    \includegraphics[width=0.5\columnwidth]{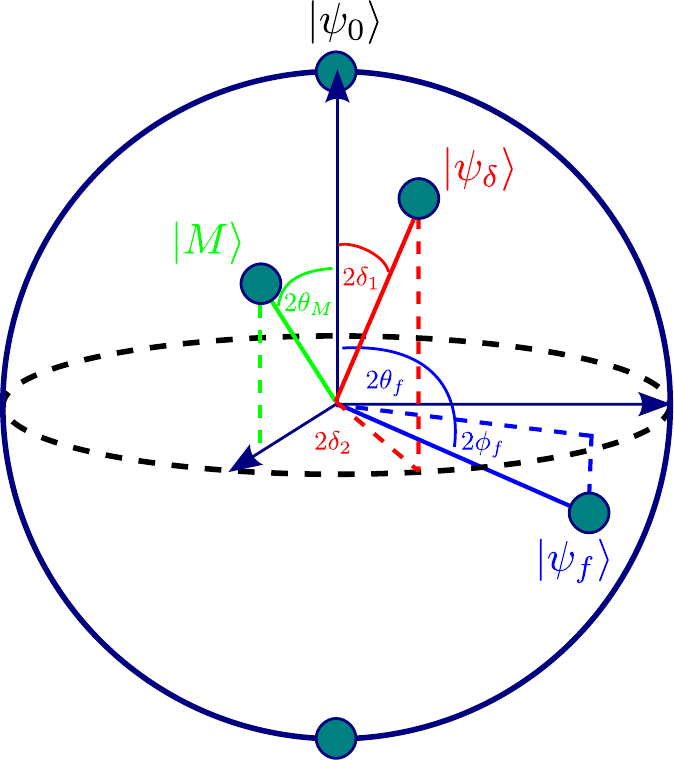}
  \end{center}
  \caption{\label{bloch2}
  A Bloch sphere visualization of the chosen gauge for the analysis of the three-component POVM case.}
\end{figure}

 Let us choose a gauge such that $\ket{\psi_0}=\ket{0}$, the measurement orientation is $\ket{M}=\cos[\theta_M] \ket{0} + \sin[\theta_M]\ket{1}$, the alternate state is $\ket{\psi_\delta}=\cos[\delta_1] \ket{0} + \sin[\delta_1]e^{i\delta_2}\ket{1}$, and the postselection orientation is $\ket{\psi_f}=\cos[\theta_f] \ket{0} + \sin[\theta_f]e^{i\phi_f}\ket{1}$ (see Fig.~\ref{bloch2}). In this gauge we obtain the following probabilities:
\begin{widetext}
\begin{align}
\bra{\psi_0}\hat{\Pi}_?\ket{\psi_0}=&\frac{1}{2} \Bigg(1-p \cos^2\left[\theta_M\right]-\cos [2\theta_f ] \left(1-p \cos [2\theta_M] \cos^2\left[\theta_M\right]-(\sqrt{1-p}-1) \sin^2[2\theta_M]\right)\nonumber\\
&+\sin [2\theta_f ] \sin [2\theta_M] \cos \left[\phi_f\right] \left(p \cos ^2\left[\theta_M\right]+\left(\sqrt{1-p}-1\right)
   \cos [2\theta_M]\right)\Bigg)\, ,\\
\bra{\psi_\delta}\hat{\Pi}_?\ket{\psi_\delta}=&\frac{1}{4}\Bigg(\left(1+\cos[2\theta_M]\cos[2\theta_f]+\sin[2\theta_M]\sin[2\theta_f]\cos\left[\phi_f\right]\right)\left(1-\cos[2\theta_M]\cos[2\delta_1]-\sin[2\theta_M]\sin[2\delta_1]\cos\left[\delta_2\right]\right)\nonumber\\
&(1-p)\left(1-\cos[2\theta_M]\cos[2\theta_f]-\sin[2\theta_M]\sin[2\theta_f]\cos\left[\phi_f\right]\right)\left(1+\cos[2\theta_M]\cos[2\delta_1]+\sin[2\theta_M]\sin[2\delta_1]\cos\left[\delta_2\right]\right)\nonumber\\
&\frac{1}{2}\sqrt{1-p}\Big(-4 \sin[\delta_1] \cos^2[\delta_1] \sin^2[\theta_f] \sin[4 \theta_M]-4 \sin [2 \delta_1] \sin [2 \theta_f] \cos^4[\theta_M] \cos[\delta_2+\phi_f]\nonumber\\
&-8 \sin^3[\theta_M] \cos[\theta_M] \left(\sin[2 \theta_f] \left(\cos ^2[\delta_1] \cos[\phi_f]-\sin^2[\delta_1] \cos[2\delta_2-\phi_f]\right)+\sin[2\delta_1] \cos[\delta_2] \cos^2[\theta_f]\right)\nonumber\\
&+8 \sin[\theta_M] \cos^3[\theta_M] \left(\sin[2 \theta_f] \left(\cos^2[\delta_1] \cos[\phi_f]-\sin^2[\delta_1] \cos[2\delta_2+\phi_f]\right)
+\sin[2\delta_1] \cos[\delta_2] \cos^2[\theta_f]\right)\nonumber\\
&+\sin[2 \delta_1] \sin[2 \theta_f]    \left(4 \sin^2[\theta_M] \cos[2 \theta_M] \cos[\delta_1-\phi_f]+    \sin^2[2 \theta_M]  \cos[\delta_2+\phi_f]\right)\nonumber\\
&-2 \cos[2 \theta_f] \sin^2[2 \theta_M] \left(-2 \sin^2[\delta_1] \cos[2 \delta_2]+\cos[2 \delta_1]+1\right)\Big)\Bigg)\, ,\\
\bra{\psi_0}\hat{\Pi}_2\ket{\psi_0}=&\frac{1}{2} \Bigg(1-p \cos^2\left[\theta_M\right]+\cos [2\theta_f] \left(1-p \cos [2\theta_M] \cos^2\left[\theta_M\right]-(\sqrt{1-p}-1) \sin^2[2\theta_M]\right)\nonumber\\
&-\sin [2\theta_f ] \sin [2\theta_M] \cos \left[\phi_f\right] \left(p \cos ^2\left[\theta_M\right]+\left(\sqrt{1-p}-1\right)
   \cos [2\theta_M]\right)\Bigg)\, ,\\  \bra{\psi_\delta}\hat{\Pi}_1\ket{\psi_\delta}=&\frac{p}{2}\left(1+\cos[2\theta_M]\cos[2\delta_1]+\sin[2\theta_M]\sin[2\delta_1]\cos\left[\delta_2\right]\right)\, .
\end{align}
\end{widetext}

%

Similar to the minimum error case, we can assume a probability distribution out of which $\ket{\psi_\delta}$ is chosen in order to deduce some optimization in the choice of the measurement directions.
We can assume, for example, that $\ket{\psi_\delta}$ is uniformly distributed on the sphere in some area around $\ket{\psi_0}$. Taking the mean over this distribution results in
\begin{widetext}
\begin{align}	
\overline{\bra{\psi_\delta}\hat{\Pi}_?\ket{\psi_\delta}}=&\frac{1}{8
   (1-\cos[2 \Delta])}\Bigg\{4-2 p+\sin[2\theta_f] \sin[2\theta_M] \cos[\phi_f] \left(2 p+(p-2) \sin^2[2 \Delta] \cos[2\theta_M]\right)\nonumber\\
&+\cos[2\theta_M] \left(\sin^2[2 \Delta] ((p-2) \cos[2 \theta_f] \cos[2 \theta_M]-p)+2 p (1-\cos [2 \Delta]) \cos[2 \theta_f]\right)\nonumber\\
&+2 \cos[2 \Delta ] (p-2-p \sin[2 \theta_f] \sin[2 \theta_M]\cos[\phi_f])+\sqrt{1-p} \Big(\frac{4}{\pi} \sin[2 \theta_f] \cos[2 \theta_M]    \sin[\phi_f] (2 \Delta -\sin[2 \Delta]
   \cos[2 \Delta])\nonumber\\
   &+\sin ^2[\Delta ] (3+\cos[2 \Delta]) \left(\sin[2\theta_f] \sin[4\theta_M] \cos[\phi_f]-2 \cos[2 \theta_f]
   \sin^2[2 \theta_M]\right)\Big)\Bigg\}\, ,\\ \overline{\bra{\psi_\delta}\hat{\Pi}_1\ket{\psi_\delta}}=&\frac{p}{2}\left(1+\cos[2\theta_M]\cos^2\left[\Delta\right]\right)\, ,
\end{align}
\end{widetext}
where $\Delta$ corresponds to the area of the probability distribution out of which $\ket{\psi_\delta}$ is selected [cf.~Eq.~\eqref{mean_std}].

\begin{figure*}
  \begin{center}
    \includegraphics[width=0.8\textwidth]{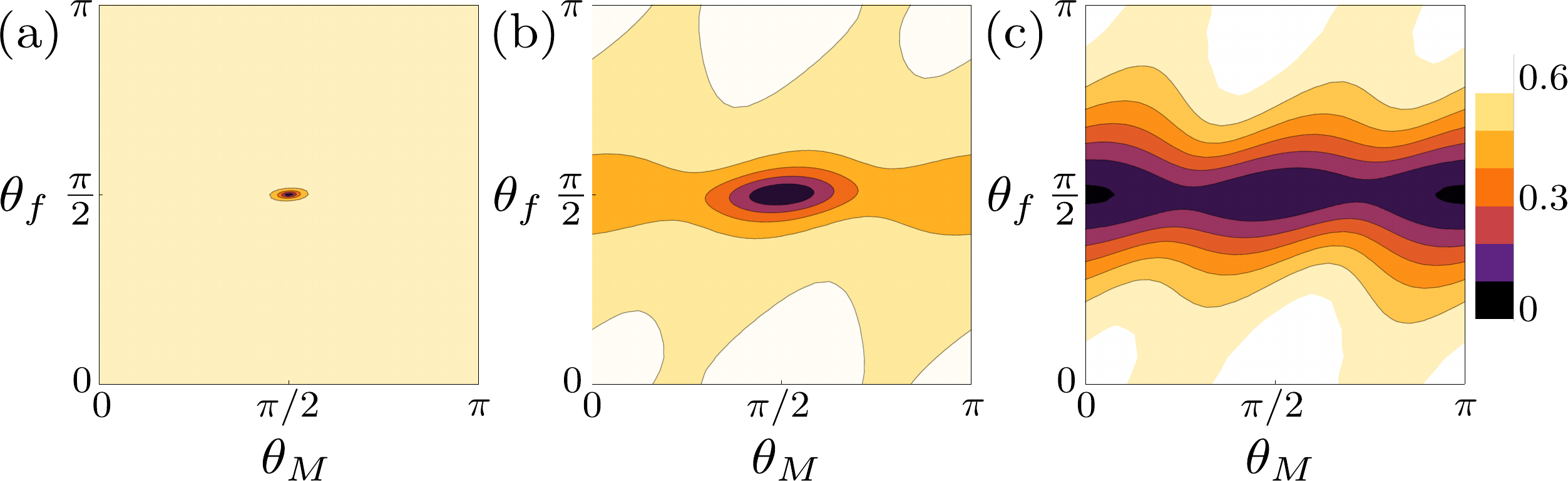}
  \end{center}
  \caption{\label{plots}
  Contour plots of the discrimination error probability, $p_{\rm err}$, obtained in the three-component POVM case [cf.~Eq.~\eqref{perr3povm}] as a function of the measurements orientations for (a) $\Delta=0.1$; (b) $\Delta=\pi/4$; (c) $\Delta=\pi$, where $\Delta$ controls the area of the probability distribution out of which $\ket{\psi_\delta}$ is selected [cf.~Eq.~\eqref{mean_std}]. In (a) and (b) the minimum error probability is obtained when both measurements are performed orthogonal to $\ket{\psi_0}$. In (c) the minimum error is obtained when the first measurement is in the direction of $\ket{\psi_0}$ and the postselection is orthogonal to $\ket{\psi_0}$. All plots are for $\phi_f=0$ and $p=0.1$.
  }
\end{figure*}

Keeping in mind the experiment discussed in the main text, we consider the specific case where both measurements are lying in the same plane ($\phi_f=0$). We can see in Fig.~\ref{plots} that the optimal measurement orientations correspond to $\theta_f = \pi/2$, i.e.~an application of $\hat{\Pi}_2$ in a direction orthogonal to $\ket{\psi_0}$ (namely, $\ket{\psi_f}=\ket{1}$). The optimal orientation of $\hat{\Pi}_1$ depends on $p$ and $\Delta$. Importantly, for discrimination between nearby states (requiring $\Delta\ll 1$), the optimal orientation of $\hat{\Pi}_1$ is nearly orthogonal to $\ket{\psi_0}$, as in the case of a single von Neumann measurement. When the unknown state can lie anywhere on the Bloch sphere ($\Delta=\pi$) the optimal orientation of $\hat{\Pi}_1$ is in the direction of $\ket{\psi_0}$ itself.

\section{IV. Multi-copy state discrimination}
\label{multi_copy}
In the main text, we introduce the SNR analysis of both the standard strong measurement approach and the null values one. Here, we provide more details of the calculations leading to Eqs.~(3) and (6) in the main text. We also consider, here, general states on the Bloch sphere, i.e., states which are not confined to the great circle of polar angles.

\subsection{A. Standard scheme (Minimum Error)}
 The statistical measure for a multi-copy discrimination is defined by Eqs.~(1)-(3) in the main text. Here, we detail the derivation of the SNR in Eq.~(3), i.e., the SNR achieved through individual standard strong measurement, $M_s$, performed on $N$ copies of a qubit. In this benchmark case, the occupation of the state $\ket{M}=\cos[\theta_M] \ket{0} + \sin[\theta_M]e^{i\phi_M}\ket{1}$ is measured. The probabilities to detect a qubit state in $\ket{M}$ in any single attempt are $P(M_{s,\delta})=|\braket{M}{\psi_\delta}|^2=\left(1+\cos [2\delta_1 ] \cos [2\theta_M ]+\sin [2\delta_1 ] \sin [2\theta_M ] \cos \left[\phi_M-\delta_2\right]\right)/2$, $P(M_{s,0})=|\braket{M}{\psi_0}|^2=\cos^2[\theta_M]$ for the states $\ket{\psi_\delta}$, $\ket{\psi_0}$, respectively. The uncertainty in the signal [cf.~Eq.~(2) in the main text] is defined by
\begin{align}
\Delta S_{\rm std}&=\sqrt{\left(\frac{\partial S}{\partial N_{s,\delta}}\right)^2\Delta N_{s,\delta}^2 +\left(\frac{\partial S}{\partial N_{s,0}}\right)^2\Delta N_{s,0}^2}\nonumber\\
&=\sqrt{N_{s,\delta}+N_{s,0}}\, ,
	\label{noise_simpsup}
\end{align}
where for the second equality we assumed Poissonian noise (which is dominant for coherent light experiments discussed below), i.e.~$\Delta N_{s,\delta}^2=N_{s,\delta}$ and $\Delta N_{s,0}^2=N_{s,0}$.

Note that the obtained SNR [cf.~Eq.~(3) in the main text] is in fact a student-T test~\cite{miller2004probability}
\begin{equation}
	\mathrm{SNR}_{\text{std}}=\frac{S}{\Delta S}\approx \sqrt{2}|\sin[\theta_M]|\delta_1 \sqrt{N} > z_{1-\eta}\, ,
	\label{SNR_simpsup}
\end{equation}
where the approximation is for $\delta \ll 1$, and $z_{1-\eta}$ is the critical value of the standard normal distribution function, $\Phi(z_{1-\eta})=\frac{1}{2}\left[1+\text{Erf}(z_{1-\eta}/\sqrt{2})\right]\equiv 1-\eta$, i.e. Eq.~\eqref{SNR_simpsup} describes the required SNR for which the signal (numerator) measured in units of  the standard deviation (denominator), can be discerned  with a given fidelity~\cite{miller2004probability}. Indeed, in this approach the maximal SNR is obtained when the measurement orientation, $\ket{M}$, is orthogonal to $\ket{\psi_0}$. This corresponds to the optimal measurement orientation obtained by the single-copy analysis (see Section \ref{single_copy}).

\subsection{B. Null Values (three-component POVM)}
In the main text, we wrote the statistical test for a multi-copy discrimination using null values [cf.~Eqs.~(4)-(7) in the main text]. Here, we provide more details on how to obtain this result and how to calculate the NVs.

In this case (cf. Fig. 1 in the main text), the qubit state is measured twice.
The first measurement, $M_w$, is a partial-collapse measurement on the states $\ket{M}$ with probability $p$.
Subsequently the qubit state is (strongly) measured a second time (postselected), $M_s$, to be in the state $\ket{\psi_f}$ (click) or $\ket{\bar{\psi}_f}$ (no click), where $\ket{\psi_f}=\cos[\theta_f] \ket{0} + \sin[\theta_f]e^{i\phi_f}\ket{1}$. We proposes to discriminate between the two possible initial qubit states via repeating the protocol for $\ket{\psi_0}$ and $\ket{\psi_\delta}$ and comparing the respective conditional outcomes of $P(M_{w,0}|\bar{M}_{s,0})$ and $P(M_{w,\delta}|\bar{M}_{s,\delta})$, i.e.~[having a click the first time conditional to \textit{not} having a click the second time]. In other words, we define our signal to be $S_{\text{NV}}\equiv P(M_{w,\delta}|\bar{M}_{s,\delta})-P(M_{w,0}|\bar{M}_{s,0})$. Note that this procedure can also be written as a correlation between outcomes of POVMs (see Section \ref{povm}).

Our protocol takes advantage of the correlation between the two measurements. To shed some light on its outcome we calculate explicitly the conditional probabilities following the measurement procedure sketched in Fig.~1 in the main text.  For example, if the first measurement results in a ``click'' the system's state is destroyed and any subsequent measurement on the system results in a null-result, implying $P(M_s|M_w)=0$, and $P(\bar{M}_s|M_w)=1$.
This represents a classical correlation between two measurements. By contrast, $P(\bar{M}_s|\bar{M}_w)$ embeds non-trivial quantum correlations.
The first partial-collapse measurement of a given preselected state $\ket{\psi_\delta}$ results in the detector clicking with probability $P(M_{w,\delta})=p|\braket{M}{\psi_\delta}|^2$.
If no click occurs [with probability $P(\bar{M}_{w,\delta})=1-P(M_{w,\delta})$], the qubit's state is modified by the measurement back action into
\begin{align}
\resizebox{.8\hsize}{!}{$\displaystyle \ket{\psi_{\delta,p}}=\frac{\big[\braket{\bar{M}}{\psi_\delta}\ket{\bar{M}}+\sqrt{1-p}\braket{M}{\psi_\delta}\ket{M}\big]}{\sqrt{P(\bar{M}_{w,\delta})}}\,.$}
\end{align}
A second strong measurement, $M_s$, yields a click [no click] with probability $P(M_{s,\delta}|\bar{M}_{w,\delta})=|\langle \psi_f|\psi_{\delta,p}\rangle |^2$   $\left[P(\bar{M}_{s,\delta}|\bar{M}_{w,\delta})=|\langle \bar{\psi}_f|\psi_{\delta,p}\rangle |^2\right]$.
Finally, using Bayes theorem, we can write 
\begin{align}
P(M_{w,\delta}|\bar{M}_{s,\delta}) &= \frac{P(M_{w,\delta})}{[ P(M_{w,\delta}) + P(\bar{M}_{w,\delta}) P(\bar{M}_{s,\delta}|\bar{M}_{w,\delta}) ]}\nonumber\\
 &=N_{w,\delta}/ (N_{w,\delta}+N_{p,\delta})\,,
\end{align}
where the last equality is obtained by taking the measured estimator for the conditional probability [Theoretically $N_{w,\delta}=N P(M_{w,\delta})$ is the number of clicks in the first measurement and $N_{p,\delta}=N P(\bar{M}_{s,\delta},\bar{M}_{w,\delta})$ is the number of no-clicks in the (second) postselection]. Note that if the detector clicks in the first measurement, the protocol is truncated, and no second step is to be carried out. This finally leads to the signal reported in Eq.~(5) in the main text,
\begin{equation}
	S_{\text{NV}}= \left|\frac{N_{w,\delta}}{ N_{w,\delta}+N_{p,\delta}}-\frac{N_{w,0}}{ N_{w,0}+N_{p,0}}\right|\,.
	\label{signal_nullwvsup}
\end{equation}
In complete analogy with the case of a single strong measurement, we define the uncertainty in the signal
\begin{align}
\Delta S_{\text{NV}}=\sqrt{\sum_{i=w,p}\sum_{j=0,\delta}\left(\frac{\partial S_{\text{NV}}}{\partial N_{i,j}}\right)^2\Delta N_{i,j}^2}\, ,
	\label{noise_nullwvsup}
\end{align}
where 
\begin{align}
\left(\frac{\partial S_{\text{NV}}}{\partial N_w}\right)^2\Delta N_w^2=&\left[\frac{1}{N_w+N_p}-\frac{N_w}{(N_w+N_p)^2}\right]^2 N_w\,\\
\left(\frac{\partial S_{\text{NV}}}{\partial N_p}\right)^2\Delta N_p^2=&\frac{N_w^2 N_p}{(N_w+N_p)^4}\,.
\end{align}
The obtained SNR [cf.~Eq.~(6) in the main text] is in fact also a student-T test
\begin{equation}
\text{SNR}_{\text{NV}}=S_{\text{NV}}/\Delta S_{\text{NV}} > z_{1-\eta}\,,
\label{nwvSnr}
\end{equation}
for discriminating between the two states. In the main text we show that it is preferable to use a tuned NV discrimination than a standard one.

\section{V. Experimental Methods} 
\label{experimental_methods}
The experimental setup is shown in Fig.~2. A helium-neon laser at 633 nm is spatially filtered with a single mode fiber and attenuated to the picowatt level. The resulting photons are passed through a fixed polarizer and then through a movable half-wave plate and polarizer;  this ensures a pure polarization state ($\ket{\psi_\delta}$) and relatively constant flux as the angle of the polarization state is varied. The power at this stage is estimated by counting photons with a single photon detector when the removable mirror is in place. We note that the coupling efficiency for each single photon detector is measured; for the SNR theory calculations in the text, we include this efficiency, losses from colored glass filters to block background, as well as the quantum efficiency of the detectors.

The removable mirror is then removed and the photons are sent to a glass window (W) that weakly reflects vertically polarized light. The reflection and transmission probabilities for 633 nm light were calculated (and verified experimentally) using the Fresnel equations. The reflected photons are counted; photons that pass through the window are projected onto a linear polarization state ($\ket{\bar{\psi}_f}$) with a second polarizer (P2) and also counted. As stated previously, each port has its own coupling efficiency and related losses. 

Photons which are reflected from the front face of the window are collected and counted as $N_w$. However, photons are also lost along the way. For example, they can be reflected from the back face of the window (with $p=0.067$), causing an additional backaction on the light which passes through the window. We thus change the definition of our conditional $P(M_{w,\delta_1}|\bar{M}_{s,\delta_1})$ to include the probability of being measured at the first measurement step [being reflected from the window's front face], conditional on [not being reflected off the back face] \emph{and} [not being being absorbed by P2], i.e.~$P(M_{w,\delta_1}|\bar{M}_{s,\delta_1}\wedge \bar{M2}_{w,\delta_1})=N_w/(N_w+N_p)$, where $\bar{M2}_{w,\delta_1}$ denotes the event of not being reflected off the back face. Notably, the measured estimator $N_w/(N_w+N_p)$, in fact, takes into account all such sources of loss.

From the recorded arrival times of the photons in each mode, we can separate photon detection events into time bins; this allows for a determination of not only the average number of photons $N_w$ and $N_p$ but also their variances as we vary the input ($\ket{\psi_\delta}$) and post-selection ($\ket{\bar{\psi}_f}$) states. To compare these values to theory, we plot the expected SNR, incorporating the efficiencies of the experiment. 
For the data included here, we count photons for approximately 5s, with time bins of 150 $\mu$s or 25 ms for the null-WV or standard schemes, respectively. For technical reasons, the photon flux for the standard scheme was higher than for the null-WV scheme; however, the time bins are chosen to ensure that each method uses an equal number of prepared photons per \emph{measurement}. We subtract dark counts, which constitute much less than 1\% of the total counts on average. The deviation of the data from the theory plot is due to a small degree of ellipticity induced by the glass window possibly arising from an unknown optical coating or stress-induced birefringence. The linear polarization model giving rise to the theory plot of Fig.~3 does not include complexity of this type. Inclusion of a phase shift is straightforward and leads to a nearly perfect fit between theory and experiment \cite{ellipse}. Consequently, we expect that elimination of this ellipticity will result in an increased SNR as suggested by the theory plot in Fig.~3. We also note that the experimental data in Fig.~3 has not been corrected for detector saturation effects. Implementing this correction is also straightforward and contributes to the nearly perfect fit mentioned above.



\begin{thebibliography}{30}
\bibitem[{rot()}]{rotated}
\bibinfo{note}{This implementation is a one-to-one mapping of the theoretical
  analysis with an overall counterclockwise rotation by the polar angle
  $\Delta_M$, i.e. the reference state is $\ket{\psi_0}=\cos[\Delta_M] \ket{0}
  - \sin[\Delta_M]\ket{1}$.}
\end{thebibliography}
\end{document}